\documentclass{aa}
\usepackage{graphicx}
\usepackage{txfonts}
\usepackage{natbib}
\begin{document}
   \title{Large scale magnetic fields in viscous resistive accretion disks} \subtitle{I. Ejection from weakly magnetized disks}
	
\titlerunning{Ejection from weakly magnetized disk}

   \author{Gareth C. Murphy \inst{1} \and Jonathan Ferreira\inst{2} \and Claudio Zanni\inst{3} }

   \offprints{G. C. Murphy}

   \institute{Dublin Institute for Advanced Studies, 31 Fitzwilliam Place, Dublin 2, Ireland\\ \email{gmurphy@cp.dias.ie} \and Laboratoire d'Astrophysique de Grenoble, CNRS, Universit\'e Joseph
             Fourier, B.P. 53, F-38041 Grenoble, France\\ \email{Jonathan.Ferreira@obs.ujf-grenoble.fr} \and INAF, Osservatorio Astronomico di Torino, Strada Osservatorio 20, Pino Torinese, Italy\\
             \email{zanni@oato.inaf.it} }

   \date{Received xx; accepted xx}

  \abstract 
% context heading (optional) 
  {} 
% aims heading (mandatory) 
  {Cold steady-state disk wind theory from near Keplerian accretion disks requires a large scale magnetic field at near equipartition strength. However the minimum magnetization has never been tested 
   with time dependent simulations. We investigate the time evolution of a Shakura-Sunyaev accretion disk threaded by a weak vertical magnetic field. The strength of the field is such that the disk 
   magnetization falls off rapidly with radius.}  
% methods heading (mandatory) 
  {Four 2.5D numerical simulations of viscous resistive accretion disk are performed using the magnetohydrodynamic code PLUTO.  In these simulations, a mean field approach is used and turbulence 
  is assumed to give rise to anomalous transport coefficients (alpha prescription). }  
% results heading (mandatory) 
  {The large scale magnetic field introduces only a small perturbation to the disk structure, with accretion driven by the dominant viscous torque. However, a super fast magnetosonic jet is observed to be launched 
  from the innermost regions and remains stationary over more than 953 Keplerian orbits. This is the longest accretion-ejection simulation ever carried out. The self-confined jet is launched from a finite radial 
  zone in the disk which remains constant over time. Ejection is made possible because the magnetization reaches unity at the disk surface, due to the steep density decrease. However, no ejection is 
  reported when the midplane magnetization becomes too small. The asymptotic jet velocity remains nevertheless too low to explain observed jets. This is because of the negligible power carried away by the jet.}  
% conclusions heading (optional), leave it empty if necessary 
  {Astrophysical disks with superheated surface layers could drive analogous outflows even if their midplane magnetization is low. Sufficient angular momentum would be extracted by the
  turbulent viscosity to allow the accretion process to continue. The magnetized outflows would be no more than byproducts, rather than a fundamental driver of accretion. However, if the midplane
  magnetization increases towards the center, a natural transition to an inner jet dominated disk could be achieved.}


   \maketitle
%
%________________________________________________________________

%%%%%%%%%%%%%%%%%%%%%%%%%%%%%%%%%%%%%%%%%%%
\section{Introduction}
%%%%%%%%%%%%%%%%%%%%%%%%%%%%%%%%%%%%%%%%%%%

Accretion disks are commonly found in young stars, active galactic nuclei, cataclysmic variables and microquasars.  In order to allow material to accrete onto a central object, it is necessary to lose
some angular momentum in an efficient way.  This is possible in a disk in one of two ways, either by radial outward transport in a disk by turbulent transport
\citep{1973A&A....24..337S,1974MNRAS.168..603L} or spiral waves \citep{1999A&A...349.1003T}, or vertical transport upwards out of the disk in a jet \citep{1982MNRAS.199..883B}.

Two extreme possible disk structures can then be identified, corresponding to each of these two processes of angular momentum removal.

The Jet Emitting Disk (hereafter JED) is threaded by a large scale magnetic field of bipolar topology driving a jet (defined here as super-fast magnetosonic flow). The dominant torque in the JED
is magnetic, due to the large braking lever arm of the jet, defined by a length scale equivalent to the Alfv\'en radius \citep{1992ApJ...394..117P}. 
The pioneering jet model by \citet{1982MNRAS.199..883B} establishes a relationship between the mass loading and the magnetic lever arm of magnetocentrifugally driven outflows. But the magnetic 
field strength was left unconstrained, so in principal any magnetization at the disk surface could drive a low-enthalpy outflow. The reason lies in the fact that an ideal MHD jet model assumes the mass
loss and does not compute it as function of the disk parameters. This was precisely the goal of semi-analytical studies done by e.g. \citet{1995A&A...295..807F} and \citet{1997A&A...319..340F}, where
the disk structure has been consistently computed: these authors showed that steady-state {\bf cold} jets can be produced only with a vertical field close to equipartition. 
A few numerical experiments tested the accretion-ejection connection in a consistent way: axisymmetric magneto-hydrodynamic (MHD) simulations of resistive
accretion disks reporting the production of self-confined, quasi-steady super-fast jets \citep{2002ApJ...581..988C, 2004ApJ...601...90C, 2007A&A...469..811Z, 2009MNRAS.400..820T}
confirmed most of the results obtained with semi-analytical models. They were however done with a large disk magnetization\footnote{The magnetization is related to the usual plasma beta by 
$\mu= 2/\beta$ in gas pressure supported disks. It is however a more general concept as s it is defined with the total pressure $P_\mathrm{gas} +P_\mathrm{rad}$.}, in the range $\mu \sim 0.2 - 1$.
The inner regions of the disk from whence jets are observed to be emitted are expected to be JED-like. In the specific case of outflows from young stars, extrapolation of slitless images of Class II jets 
have constrained the launching region to be confined to a zone of radial extent $\sim$ a few AU close to the centre of the disk \citep{2004ApJ...609..261H,2007LNP...723...21C}.

The outer regions are expected to behave more like the well studied standard accretion disk, hereafter SAD \citep{1973A&A....24..337S,2002apa..book.....F}. Here the characteristic lengthscale over
which the viscous torque is exerted is of the order of $\sim \alpha_\mathrm{v} h$, where $\alpha_\mathrm{v}$ measures the level of the turbulence. Such a turbulence is assumed to arise from the development of magnetic
instabilities that are triggered in the disk whenever a magnetic field is present \citep{1991ApJ...376..214B}. This field must however be below equipartition strength (namely $B_z^2/\mu_0 \ll P$) to
avoid the stabilizing effect of the magnetic tension. Therefore, the high magnetization required by a JED impedes the development of disk turbulence which is however required to support
a steady launching: that would leave only a very tiny parameter space for stationary ejection to take place.
The SAD-JED structure has been put forward in several papers, e.g. \citet{2006A&A...447..813F,2008A&A...479..481C,2008NewAR..52...42F}.

The study of low magnetization accretion regimes has been attempted making use of fully 3D global simulations of accretion disks, threaded by a weak large scale magnetic field.
MRI sets in and accretion is quickly established \citep{2002ApJ...573..738H}. The remarkable result is that outflows are also produced \citep{2003ApJ...592.1042I}, 
especially when the imposed field is of bipolar topology \citep{Beckwith:2009ss}.  
However, many questions remain open: What controls the mass loss in these simulations? Will the outflowing plasma become a self-confined jet? Is grid resolution enough to properly describe the turbulent 
cascade? The fact is that it is still impossible to properly follow turbulence while solving for the long term evolution of large scale systems.

As a consequence, the question of super-fast magnetosonic jet formation from weakly magnetized disks is still open. In this paper we address this issue using 2.5D numerical MHD simulations
based on a mean field approximation. We explore the accretion-ejection processes from a quasi-standard accretion disk where the magnetization is very low (smaller than $10^{-3}$).
Since the magnetic field is low, we assume that turbulence triggered by the MRI is indeed present but that it provides mainly anomalous transport coefficients: a viscosity $\nu_\mathrm{v}$ and a magnetic 
diffusivity $\nu_\mathrm{m}$. On the other hand, we do not expect to observe any MRI feature (such as channel flows for instance) in our simulation because of the presence of explicit viscosity and magnetic 
diffusivity effects. While measurements of the turbulent viscosity in MRI induced turbulence have been extensively reported in the literature, it is only very recently that such a work has been done for the 
turbulent magnetic diffusivity \citep{Lesur:2009bf, Guan:2009gd}. In particular \citet{Lesur:2009bf} showed that the turbulent magnetic diffusion scales like a resistivity tensor with dominant diagonal 
terms. Also, as a first approximation, an isotropic value can be safely used. Finally, the effective Prandtl number $ \mathcal{P}_\mathrm{m} = \nu_\mathrm{v}/\nu_\mathrm{m}$, given by the ratio of turbulent viscosity and diffusivity, 
has been found to be of order unity.
The mean field approximation has been successfully employed in a number of semi-analytical \citep[e.g.][]{1995A&A...295..807F, 1995ApJ...444..848L, 2000A&A...353.1115C, 2001ApJ...553..158O, 
2008ApJ...677.1221R} and numerical applications \citep[e.g.][]{2002ApJ...581..988C, 2003ApJ...589..397K, 2003A&A...398..825V, 2006A&A...460....1M, 2007A&A...469..811Z, 2009MNRAS.399.1802R} 
related to the study of magnetized accretion-ejection flows. Beside having a precise control of the diffusive and transport phenomena, the numerical experiments based on this approach provide laminar flow 
solutions which can be compared to semi-analytical models.

In section 2, we describe the numerical method used, the boundary and initial conditions. Section 3 is devoted to the description and discussion of the results obtained. Surprisingly,
super-fast jets are indeed obtained from a finite disk region and remain stable for a time span never previously achieved in the literature. Section 4 summarizes our findings and, in a companion paper
(Murphy et al., in prep), we will examine the long standing issue of the magnetic field redistribution within the disk on long (accretion) time scales.

%%%%%%%%%%%%%%%%%%%%%%%%%%%%%%%%%%%%%%%%%%%
\section{Numerical method}
%%%%%%%%%%%%%%%%%%%%%%%%%%%%%%%%%%%%%%%%%%%
\label{NumericalMethod}

%\subsection{The PLUTO MHD code}
The full visco-resistive MHD equations in axial symmetry are evolved in time using the publicly available numerical code PLUTO \citep{2007ApJS..170..228M}.  
The solved equations are: the continuity equation

\begin{equation}
\frac{\partial \rho}{\partial t}+\nabla\cdot(\rho \vec u)=0 \; ;
\end{equation}

the conservation of momentum equation
\begin{equation}
\frac{\partial}{\partial t}\left( \rho \vec u \right) +\nabla \cdot \left[ \rho \vec u \otimes \vec u +\left( P^* \right) {\overline {\overline{\vec I} }} +\vec{B} \otimes \vec{B} +
{\overline {\overline{\vec T}}} \right] + \rho \vec \nabla \Phi_\mathrm{G} =0 \; ;
\end{equation}

the induction equation
\begin{equation}
\frac{\partial \vec{B} }{\partial t}+ \nabla \times ( \vec{B} \times \vec{u} + \nu_\mathrm{m} \vec{J} ) = 0 \; ;
\end{equation}

the conservation of energy equation
\begin{equation}
\frac{\partial E}{\partial t} +\nabla \cdot \left[ \left( E + P^*\right) \vec{u} - (\vec{u}\cdot\vec{B})\vec{B} + \nu_\mathrm{m} \vec{J} \times \vec{B} - \vec{u} \cdot {\overline
{\overline{\vec T}}} \right] = S \; ,
\end{equation}
where $S=-\rho\vec{u} \nabla \Phi_\mathrm{G} + L_\mathrm{c}$, and $L_\mathrm{c}$ is the local cooling term (see below). The total energy density is defined as
\begin{equation}
E = \frac{1}{2}\rho |\vec{u}|^2 + \frac{P}{\gamma-1}+ \frac{1}{2} |\vec{B}|^2 \; ,
\end{equation}
and the total pressure (thermal and magnetic) is
\begin{equation}
P^* = P + \frac{1}{2}|\vec{B}|^2 \; .
\end{equation}

The equations are written and solved in dimensionless form, thus without $\mu_0$ coefficients. The equation of state is the ideal gas equation. 
Here, $\rho$ is the mass density, $\vec{u}$ the velocity, $P$ the gas pressure, $\vec{B}$ the magnetic field, $\Phi_\mathrm{G} = - GM / \sqrt{r^2
+ z^2} $ is the gravitational potential of the central mass, $\vec{J}=\nabla \times \vec{B}$ is the current density,$\nu_\mathrm{m}$ the magnetic diffusivity and 
$\gamma = 5/3$ is the ratio of specific heats.  
The viscous stress tensor ${\overline {\overline {\vec T}}}$ is defined as
\begin{equation}
{\overline {\overline {\vec T}}} = \eta_\mathrm{v} \left[ \left(\nabla  \vec{u}\right) + \left(\nabla  \vec{u}\right)^T - \frac{2}{3}\left(\nabla\cdot\vec{u}\right) \vec{I} \right] \; ,
\end{equation}
where $\eta_\mathrm{v}$ is the dynamic viscosity. See Appendix \ref{AddNumCond} for the expression of the tensor components. As is customary, the kinematic viscosity 
is defined as $\nu_\mathrm{v} = \eta_\mathrm{v}/\rho$.

As stressed above, we follow a mean field approach where the turbulence is crudely modeled by mere transport coefficients: a viscosity $\nu_\mathrm{v}$ and a magnetic diffusivity $\nu_\mathrm{m}$.  
Consistently with this approximation, a \citet{1973A&A....24..337S} alpha prescription is then employed. This assumes that the viscosity is proportional to the heightscale of
the disk, $h$, and some characteristic velocity, in this case the sound speed, $c_\mathrm{s}$, namely
\begin{equation}
\alpha_\mathrm{v} \equiv \frac{3}{2}\frac{\nu_\mathrm{v}}{c_\mathrm{s} h} \; .
\end{equation}
We assume that the disk is not flat, but will have initially a constant aspect ratio $\varepsilon= h/r= c_\mathrm{s}/V_\mathrm{K}= 0.1$.
As an initial condition for the alpha accretion disk, we take the perturbative solution of the steady-state disk equations found in \citet{2009A&A...508.1117Z} and the references therein. The disk is in hydrostatic equilibrium 
and accretion is driven by the viscous stress tensor alone.  This particular solution provides reasonable vertical and radial profiles of all quantities that are suitable for a SAD (see Appendix
\ref{AddNumCond} for more details). 
A not so well known bias of the alpha prescription in 2D flows is that, below a critical value found to be $\alpha_\mathrm{crit} \sim 0.685$, there is a backflow on the
disk midplane \citep{1984SvA....28...50U}. This is certainly unphysical and arises only from the functional form of the stress tensor used to mimic turbulence. In order to circumvent this bias, we
used $\alpha_\mathrm{v}=0.9$\footnote{This value might be seen too large when compared to the small mean field used in the disk.
However, note that the main effect of a large $\alpha_\mathrm{v}$ is to reduce the accretion time scale, while still maintaining it well below the Keplerian one.}.

Consistently with the recent \citet{Lesur:2009bf} results, we assume that the effective magnetic Prandtl number $\mathcal{P}_\mathrm{m} = \nu_\mathrm{v}/\nu_\mathrm{m}$ is of
order unity: for simplicity we set $\mathcal{P}_\mathrm{m} = 2/3$ in all simulations. Again, we stress that this is a strong simplification of highly complex phenomena but also an
unavoidable price to pay if one seeks for long term evolution of global systems, such as accretion disks and their related jets. 
With a constant $\mathcal{P}_\mathrm{m}$, the viscosity and resistivity will follow the same radial and vertical profiles. They decrease smoothly with height until they become negligible, allowing a transition to
a magnetized ``corona'' in ideal MHD regime. Since MRI induced turbulence is quenched when the magnetic field becomes close to equipartition \citep{1991ApJ...376..214B}, there will be a height where
the accretion flow cannot be turbulent anymore \citep{2008ApJ...677.1221R}. For simplicity, we assume that it corresponds to the disk surface (see Appendix \ref{AddNumCond}). We thus follow
\citet{2002ApJ...581..988C,2004ApJ...601...90C} and \citet{2007A&A...469..811Z} in neglecting the turbulent viscosity and turbulent resistivity in the highly magnetised corona.

 In a real accretion disk, the local heating due to turbulence (here crudely modeled by alpha prescriptions for resistivity and viscosity) would be balanced by both turbulent transport and
radiative cooling. While the former cooling term needs full 3D calculations, the latter can be done in 2D but requires radiative transfer. Both effects are far beyond the scope of the present
work. Hence, by including a "cooling" function $L_\mathrm{c}$ such that
\begin{equation}
L_\mathrm{c} = \nu_\mathrm{m} J^2 + \frac{1}{2\eta_\mathrm{v}} \left[ T_{rr}^2+T_{zz}^2+T_{\phi\phi}^2+ 2 (T_{rz}^2+T_{r\phi}^2+T_{z\phi}^2 ) \right] \; ,
\end{equation}
we can exactly balance both resistive and viscous heating terms. Our disk evolution is therefore adiabatic despite the presence of transport coefficients within the disk. This is certainly a caveat,
shared by most today MHD simulations, and deserves further investigation. On the other hand, it allows to avoid in a simple way the otherwise unavoidable (and unphysical) vertical expansion of the
disk when heating is present without any kind of cooling.
A static atmosphere in pressure equilibrium is set above the disk and a large scale magnetic field is superimposed in the whole domain.

To set the initial magnetic field, we use the magnetic flux function $\Psi$ such that $\vec{B} = \nabla \Psi \times \vec e_\phi/r$. We take the particular form
\begin{equation}
\Psi (r,z) = 4 B_0 r_0^2 \left(\frac{r}{r_0}\right)^{1/4} \frac{m^{7/4}}{\left( m^2 + z^2/r^2\right)^{7/8}} \; ,
\end{equation}
where $B_0 = \sqrt{\mu_0 \mu(r_0) P_\mathrm{d0}}$, $P_\mathrm{d0}$ is the thermal pressure of the disk and $\mu$ is the disk magnetization at the disk midplane of the inner radius $r_0$. The parameter $m$ describes
the initial bending of the magnetic field lines, and is set to $0.935$ in all simulations. This leads to a magnetic field such that the disk magnetization $\mu$ varies initially as $r^{-1}$, starting
from $\mu(r_0) = 2\times 10^{-3}$ (one simulation is done with $\mu(r_0) = 2\times 10^{-4}$). Such a small value for the large scale field is chosen to ensure that it is only a tiny perturbation to
the initial SAD structure on the midplane.

The MHD system of equations has been solved numerically exploiting the MHD module provided with PLUTO. The code has been
configured to perform second-order piecewise linear reconstruction of primitive variables, with a Van Leer limiter for the density and magnetic field components 
and a minmod limiter for the thermal pressure and velocity components. To compute the intercell fluxes, a HLL Riemann solver has been employed \citep{harten:35}, 
while second order in time has been achieved using a Runge-Kutta scheme. 
The solenoidal condition, $\vec\nabla\cdot\vec B =0$, is preserved using the constrained transport method \citep{1988ApJ...332..659E}.
The viscous and resistive terms have been treated explicitly, using a second-order finite difference approximation for the dissipative fluxes and checking the 
diffusive timestep. 

A uniform resolution grid of 512 by 1536 cells is used. This describes a domain of 40$r_0$ by 120$r_0$, where $r_0$ is the inner radius of the disk. An outer, stretched grid is extended for a further
512 cells in the radial direction and 1536 in the vertical direction, thus describing in total a region 280$r_0$ by 840$r_0$.  To examine the effects of a higher resolution (see Section
\ref{NumericalResolution}), the same grid is used again but this time only describing a region of 5$r_0$ by 15$r_0$, with a logarithmically stretched grid outside this region 35$r_0$ by 105$r_0$.  A
disadvantage of higher resolution is that the number of timesteps to reach accretion timescales becomes prohibitive.

%\subsection{Boundary conditions}
The boundary conditions are axial symmetry on the rotation axis and equatorial symmetry for the disk midplane. The upper $r$ and $z$ boundaries border on a logarithmically stretched grid which ensures
that the magnetized outflow never reaches the boundaries.  The ghost cells at the upper $r$ and $z$ boundaries are set to equal the values inside the domain (the numerical approximation to an
``outflow'' boundary condition).  The gravitational potential has a singularity at the origin, so a rectangular portion of the simulation close to the origin is excluded, as in
\citet{2002ApJ...581..988C}.  The right boundary of the rectangular region is a sink and the upper boundary injects a small amount of material into the grid at the escape velocity (with density 1.1
times the local initial density). This keeps the axis sufficiently dense to ensure that unphysically low densities are not produced on the axis by the Lorentz force. For the poloidal magnetic field
the boundary condition can be expressed in terms of the toroidal electric field, $E_\phi$. Assuming flux is not advected into the central object, we impose $E_\phi=0$.

Throughout the paper the distances are expressed in units of $r_0$, which is the inner disk radius in the simulation. Velocities have been normalized on the Keplerian speed at $r_0$,
$V_\mathrm{K,0}=\sqrt{GM/r_0}$. Densities are expressed in units of $\rho_\mathrm{d0}$, the initial disk density at its inner radius. The times are in units of the Keplerian orbital period $\tau_\mathrm{K0}= 2\pi
r_0/V_\mathrm{K,0}$.  Pressures are given in units of $\rho_\mathrm{d0}V_\mathrm{K,0}^2$ while the magnetic field is expressed in units of $\sqrt{\mu_0 \rho_\mathrm{d0}V_\mathrm{K,0}^2}$.

 For ease of reproducibility, the C subroutines defining initial conditions and boundary conditions are available from the authors on request. The numerical code PLUTO is publicly available from
the URL http://plutocode.to.astro.it.

%%%%%%%%%%%%%%%%%%%%%%%%%%%%%%%%%%%%%%%%%%%%%%
\section{Ejection from weakly magnetized disks}
\label{Ejection}
%%%%%%%%%%%%%%%%%%%%%%%%%%%%%%%%%%%%%%%%%%%%%%

\subsection{Global description}
\begin{figure*}[t]
\sidecaption
   \includegraphics[width=12cm]{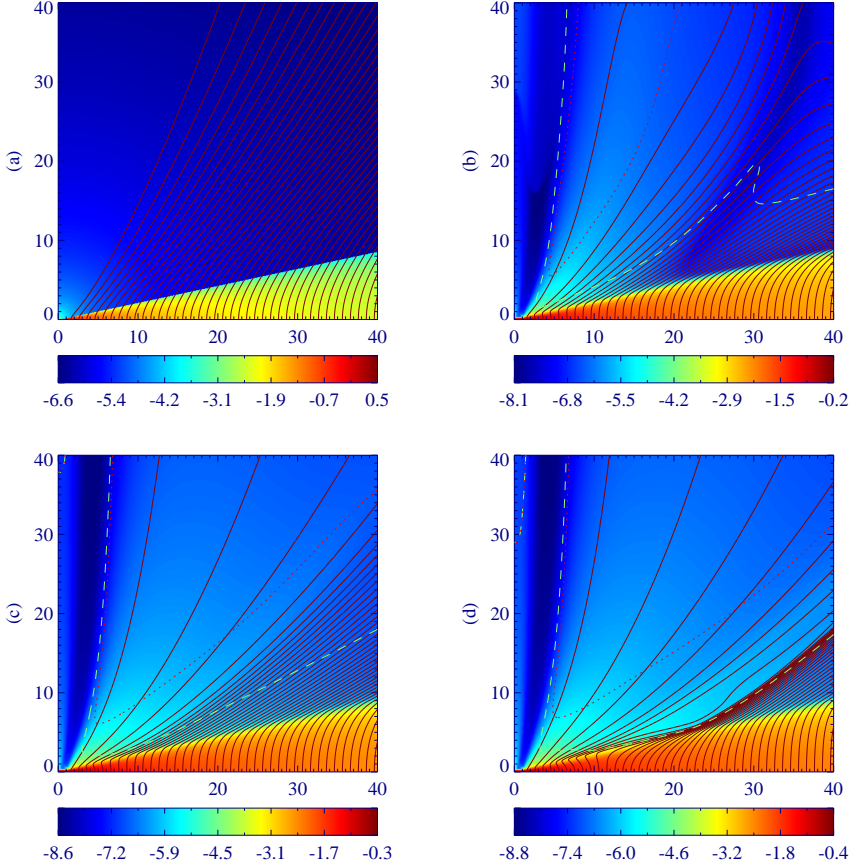}
   \caption{Log of mass density at times (a) $t=0 \tau_\mathrm{K0}$, (b) $t=31 \tau_\mathrm{K0}$, (c) $t=160 \tau_\mathrm{K0}$, (d) $t=953 \tau_\mathrm{K0}$. The fast surface and the Alfv\'en surface are overplotted with
   dotted and dashed lines respectively. The super fast-magnetosonic outflow (the jet) is launched only from a small inner region, located between $r=1$ and $r=5$.  The extension of this zone remains constant over time.}
   \label{bigfig}%
\end{figure*}

%%%%%%%%%%%%%%%
When the simulations starts in the first visible phenomenon is the triggering of the familiar vertical torsional Alfv\'en wave \citep{1980ApJ...237..877M,1997ApJ...482..712O}.  It is due to the
differential rotation between the Keplerian disk and the initially non rotating atmosphere. But after a few inner disk rotations, a proper MHD outflow is launched from the disk, developing a bow shock
and compressing the ambient material and the preceding torsional flow. Figure \ref{bigfig} shows a plot of the jet density in the poloidal plane together with the fast and Alfv\'en surfaces and
magnetic field lines. A superfast jet is launched within a relatively narrow region at the disk surface up to $r=5$. Matter launched from this region crosses the slow and Alfv\'en surfaces close to
the disk surface and is accelerated up to the fast magnetosonic surface.

\begin{figure}
   \centering 
   \includegraphics[width=\columnwidth]{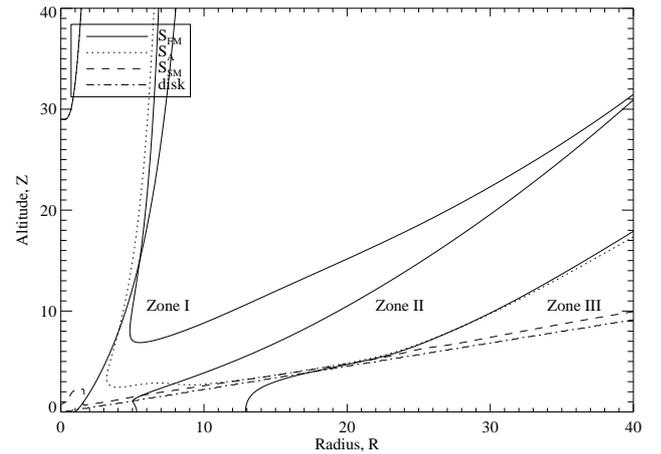}
   \caption{Poloidal cross section showing various zones at a time t=953$\tau_\mathrm{K0}$: the critical surfaces of the MHD outflow (slow magnetosonic $S_\mathrm{SM}$, Alfv\' en $S_\mathrm{A}$, fast magnetosonic $S_\mathrm{FM}$) and the disk
   surface. Field lines anchored at $r=1,5.1, 13$ are also shown, delimiting the three zones (see text)}.
              \label{surfaces}%
    \end{figure}

%%%% definition of h, disk-jet zones
Along the $z$ direction, the numerical simulation can be characterised as divided into two main zones, a resistive zone, where resistive effects are important (the disk), and an ideal MHD zone, 
where ideal MHD is strictly enforced (the jet and atmosphere).  The disk surface could be defined as the altitude where all transport coefficients vanish. We choose rather to define the disk surface 
as the altitude where the radial velocity component vanishes, marking therefore a clear transition between underlying accreting layers ($u_r<0$) and outflowing upper layers ($u_r>0$). 
Figure \ref{surfaces} shows these various surfaces at the final time $t=953 \tau_\mathrm{K0}$. Note however that they do not evolve much over time as it can be seen in Fig. \ref{bigfig}.

This result is unexpected since the normal requirement for a steady cold MHD disk wind is a near equipartition midplane magnetization \citep{1995A&A...295..807F}.  A second surprising feature is that
the outflow is launched from a clearly defined region centrally located in the disk that does not evolve from the entire disk region, even after 953 disk rotations.  This is in contrast to the
previous results of \citet{2007A&A...469..811Z} where, as the simulation evolves in time, the outflowing region moves outward on the Keplerian timescale.  In fact, the global accretion-ejection
configuration exhibits three distinct zones in the the radial direction. Zone I corresponds to the innermost radii where anchored field lines give rise to a super-fast jet, namely from $r=1$ to $r= 5$. 
Then an intermediate zone gives birth to a sub-fast but still super-Alfv\'enic outflow. Zone II goes from $r= 5$ to $r= 13$. Such a zone is expected to be unsteady as any FM wave can travel upstream. 
Finally, the last zone goes from $r=13$ up to the outermost radius and corresponds to negligible outflowing material that remains always sub-Alfv\'enic.

\begin{figure}
   \centering 
   \includegraphics[width=\columnwidth]{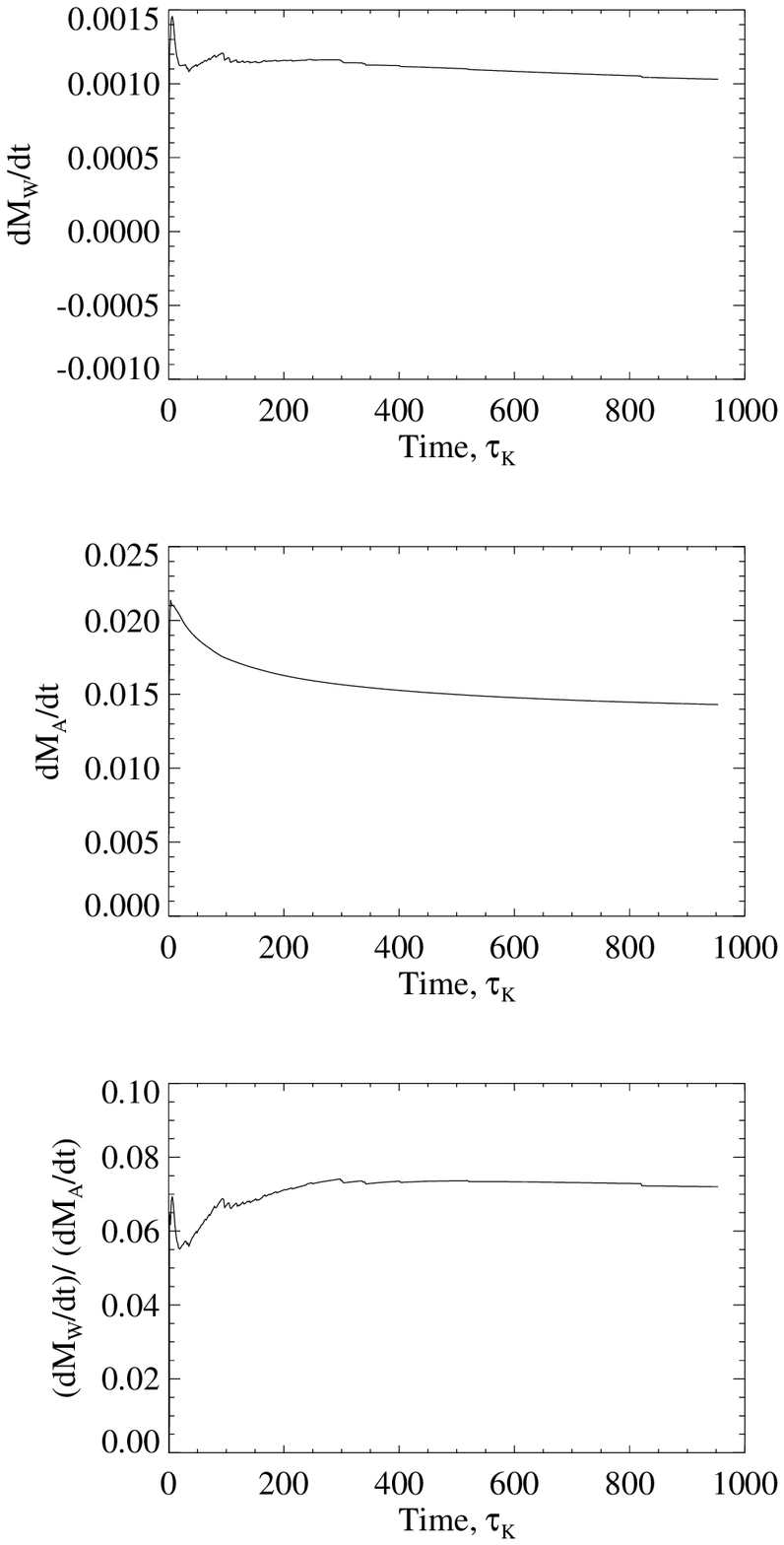}
   \caption{Time evolution of the ejection rate (top), calculated by integrating the mass flux over the super-fast region only, half disk accretion rate (middle) measured at the inner radius and the
	ejection to accretion rate ratio (down). After an initial transient phase that lasted up to 200 $\tau_\mathrm{K0}$, the global system reached a quasi steady-state.}
              \label{acc_rate}%
    \end{figure}

%% mass loss
This is the longest accretion disk simulation ever done so far (953 inner periods) where the jet remains steady. Figure \ref{acc_rate} shows the accretion (measured on one half disk thickness) and
 ejection (in one jet) rates as well as their ratio plotted over time.  $\dot M_\mathrm{w}$ is computed at the disk surface, defined by the height where the poloidal velocity reaches zero, and from a radius
 $r=1.4$ to $r=5.0$. As can be seen from the figure, the ejection to accretion mass loss is approximately 7\%. While smaller than that obtained for a disk with a larger magnetization, this
 mass loss is by no means negligible.
We also inject a small amount of mass $\dot M_\mathrm{inj}$ at the surface of the internal boundary which is of the order of 1\% of $\dot M_\mathrm{w}$ and thus negligible when compared to $\dot M_\mathrm{w}$.

%Global energy budget
 The accretion power is computed as the difference between the flux of mostly mechanical energy $E= \frac{u^2}{2} + \frac{\gamma}{\gamma-1}\frac{P}{\rho} + \Phi_\mathrm{G} $ entering the disk at its outer edge
and leaving it at its inner edge, namely
\begin{equation}
P_\mathrm{acc} = \int_\mathrm{out} \rho E \vec{u}_\mathrm{p} \cdot d\vec{S} - \int_\mathrm{in} \rho E \vec{u}_\mathrm{p} \cdot d\vec{S} \; ,
\end{equation}
where the integration is performed on a vertical section of the disk. The jet power is calculated as the sum of all energy fluxes (mechanical and Poynting) leaving the disk,
\begin{equation}
P_\mathrm{jet}=P_\mathrm{mech,jet}+P_\mathrm{MHD,jet} \; ,
\end{equation}
where
\begin{equation}
P_\mathrm{mech,jet}= \int_\mathrm{S} \rho E \vec{u}_\mathrm{p} \cdot d\vec{S} 
\end{equation}
\begin{equation}
P_\mathrm{MHD,jet}= \int_\mathrm{S} \vec{E} \times \vec{B} \cdot d\vec{S} \; .
\end{equation}
Here, the integration has been made in a control volume defined by the inner radius $r_\mathrm{in} = 1.4$ and an outer radius $r_\mathrm{out} =  5.0$.

The theoretical global energy budget should then be
\begin{equation}
P_\mathrm{acc} + P_\mathrm{visc}= P_\mathrm{jet} + P_\mathrm{rad} \; ,
\end{equation}
where $P_\mathrm{rad}$ is the power released into heat by both viscous and Joule terms (and is eventually radiated at the disk surface) and $P_\mathrm{visc} = \int_\mathrm{in} ( \vec u \cdot {\overline {\overline {\vec T}}})
\cdot dS$ is the influx of energy at the inner radius due to viscosity (the flux at the outer radius is negligible). In the standard accretion disk theory, such a flux of energy is set to exactly zero
through the ``zero torque condition'' at the inner boundary. This was not implemented in our simulation so that the actual power that is available (liberated) within our simulation box is $P_\mathrm{lib} =
P_\mathrm{acc} + P_\mathrm{visc}$. Also, in practice, our simulation does not include radiation but the Joule and viscous heating terms are balanced exactly by the cooling term $L_\mathrm{c}$.
Table \ref{tab:1} shows the different calculated powers. $P_\mathrm{visc}$ represents only 20\% of all the liberated power ($P_\mathrm{visc}/P_\mathrm{acc}= 0.25$): while not strictly negligible, it is only a small
fraction. In the following, we will thus compare the jet power only with the accretion power. The total jet power (MHD + mechanical) represents only 15\% of the the accretion power. In terms of
released power, the disk behaves therefore exactly like a standard accretion disk, the jet being a mere epiphenomenon. The obvious reason for that is the very weak magnetic field, unable to extract a
significant fraction of the available power (see below). Note also that the jet power is dominated by the MHD Poynting flux.

   \begin{table}
      \caption{Viscous, accretion, mechanical, kinetic powers and MHD Poynting flux.}
         \label{tab:1}
     $$
         \begin{tabular}{l l l}
            \hline \noalign{\smallskip} Power & Value \\ \noalign{\smallskip} \hline \noalign{\smallskip}
            Viscous Power $ , P_\mathrm{visc}$ & 0.000936 \\ 
            Accretion Power, $P_\mathrm{acc}$ & 0.00362 \\  
            Mechanical Power, $P_\mathrm{mech,jet}$ & -0.000183 \\ 
            Kinetic Power, $P_\mathrm{kinetic,jet}$ & 0.000197 \\ 
            Poynting Flux, $P_\mathrm{MHD, jet}$ & 0.000563 \\ 
            \noalign{\smallskip} \hline
         \end{tabular}
     $$
   \end{table}

\begin{figure}
   \centering \includegraphics[width=\columnwidth]{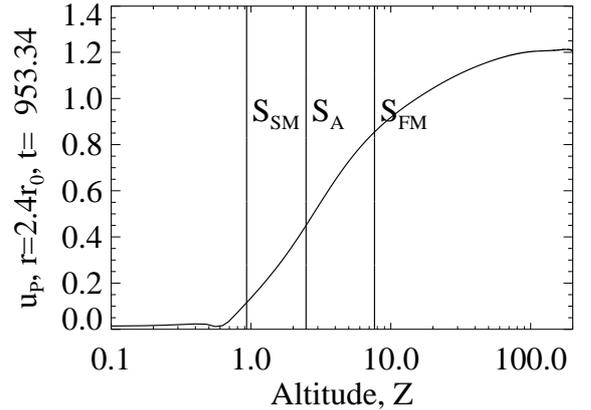}
   \caption{ Poloidal velocity normalized to the Keplerian velocity at the footpoint along a magnetic field surface.  The time is 953.3 $\tau_\mathrm{K0}$.  The radius of the footpoint of the magnetic
	surface is 2.4.  }
              \label{fig:vpol}%
    \end{figure}

%Asymptotic jet Velocity
The jet becomes super-SM very soon, almost at the disk surface and reaches the Alfv\'en speed at an altitude $z_\mathrm{A}$ significantly smaller than the corresponding Alfv\'en radius $r_\mathrm{A}$ (Fig.
\ref{surfaces}). This is again in strong contrast with self-similar {\bf cold} jet solutions where $z_\mathrm{A} \sim r_\mathrm{A}$ \citep{1997A&A...319..340F}. The flow then reaches its asymptotic velocity soon after
the fast magnetosonic surface, which is a maximum of about $\sim 1.2$ times the Keplerian value at the disk midplane (see Fig. \ref{fig:vpol}).  Thus the type of jet produced here cannot be
responsible for very high velocity Herbig-Haro jets for example. The maximum asymptotic velocity of a cold super-Alfv\'enic outflow anchored at $r_0$ is 
$u_{\mathrm{p},\infty}= V_\mathrm{K,0} \sqrt{2\lambda - 3}$, where $\lambda \simeq (r_\mathrm{A}/r_0)^2$ is the magnetic lever arm parameter. It is possible to estimate $\lambda$ using the ratio:
\begin{equation}
\frac{F_\mathrm{Poynting}}{F_\mathrm{kinetic}}= \frac{ \int_\mathrm{S} E \times B \cdot dS}{ \int_\mathrm{S} \rho u^2 u_\mathrm{p}/2 \cdot dS} = 2(\lambda-1) \; .
\end{equation}
In our case we derive a $\lambda$ of 2.4, which would provide an asymptotic velocity of 1.4 $V_\mathrm{K,0}$. This is larger than the value of 1.2 found, which hints to the fact that the magnetic structure
retained a fraction of its energy.  By computing the magnetic contribution of the total jet power further out, at $z=100$, we found that it still represents 33\% (Fig. \ref{bernoulli}). This is
again in contrast to self-similar models where all the MHD power is converted into jet kinetic power. This aspect will be discussed in a companion paper.

\subsection{The SAD structure}
%%%%%%%%%%%%%%%%%%

Despite the presence of an outflow (be it a super-FM self-confined jet or only super-A flow), the disk structure strongly resembles that of a standard accretion disk.

\begin{figure}
   \centering \includegraphics[width=\columnwidth]{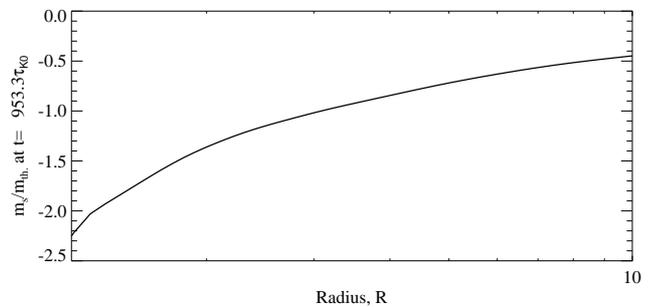}
   \caption{Ratio of midplane radial accretion sonic Mach number $m_\mathrm{s}$ at time = 953 $\tau_\mathrm{K0}$ to its theoretical value, $\alpha_\mathrm{v} c_\mathrm{s} /V_\mathrm{K}$. Accretion clearly proceeds at a rate controlled by the
		anomalous viscosity, the jet torque being negligible. }
              \label{fig:mach_acc}%
    \end{figure}

%Accretion Mach number, torque and jet power
Most of the released power is radiated away: this implies that the main dominant torque is the viscous one. Accretion proceeds therefore throughout the disk with a Reynolds number of order unity. The
value of the accretion Mach number $m_\mathrm{s}= -u_r/c_\mathrm{s}$ at $z=0$ is a good test as its fiducial value in a SAD should be of the order of $m_\mathrm{th} = \alpha_\mathrm{v} h/r$ \citep{1994ApJ...423..736R}. Figure
\ref{fig:mach_acc} shows the radial profile of $m_\mathrm{s}/(\alpha_\mathrm{v} h/r)$ at a time = 953 $\tau_\mathrm{K0}$.  The theoretical approximation is very close to the simulated one, up to a factor of about two. This is
very reasonable given the fact that the actual expression of the torque involves radial derivatives. The action of the jet is therefore totally negligible on midplane accretion.

Now, let $\Lambda$ be the ratio of the magnetic (jet) to the viscous torque averaged over the disk thickness. Since most of the available power is stored into rotation in a thin accretion disk, the
fact that the jet carries a tiny fraction of the accretion power is directly related to a negligible torque on the bulk of the disk mass. Analytically, this may be written
\begin{equation}
\frac{P_\mathrm{jet}}{P_\mathrm{acc}} \simeq \frac{\Lambda}{1 + \Lambda} \frac{- B_\phi^+}{B_z} \sim \Lambda \; .
\end{equation}
At each radius within zone I, we vertically integrate the torques and obtain thereby a distribution $\Lambda(r)$. It is relatively smooth, with small deviations from an average value of approximately
0.15. This is consistent with the ratio $P_\mathrm{MHD,jet}/P_\mathrm{acc}$ of 0.155 computed from Table \ref{tab:1}.

\begin{figure}
   \centering \includegraphics[width=\columnwidth]{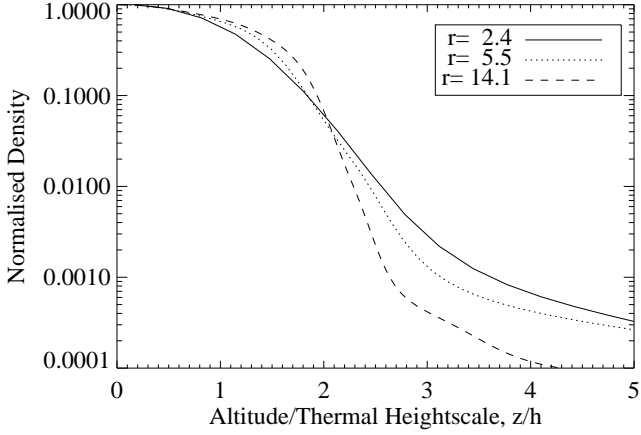}
   \caption{Vertical density profiles at $r=2.4,5.5$, and $r=14.1$, normalized to their midplane value. At the inner radius, the initial steep gradient has been flattened out. This is a signature of
   mass loss from the disk. The time is at 953 $\tau_\mathrm{K0}$.}
              \label{denlaunchreg}%
    \end{figure}

%%%%%%%% density profile
Although the action of the jet is negligible on the equatorial accretion motion, it has some impact on the disk vertical structure. This is illustrated for instance in Fig. \ref{denlaunchreg} where
the density profile is shown at different radii. The profiles have been normalized to the midplane density and plotted against $z/h(r)$ where $h(r)$ is the local thermal heightscale. Clearly, the
profile located at $r=2.4$, that is within the ejecting region, becomes flatter than that from $r=14.1$ (outside the ejecting zone), has a result of the mass loss. Notice also the dramatic decrease in
density of about four decades at the disk surface. We shall come back to this feature later on.

\subsection{Electric currents}
%%%%%%%% field and currents @ zones 1, 2 and 3

\begin{figure}
   \centering \includegraphics[width=\columnwidth]{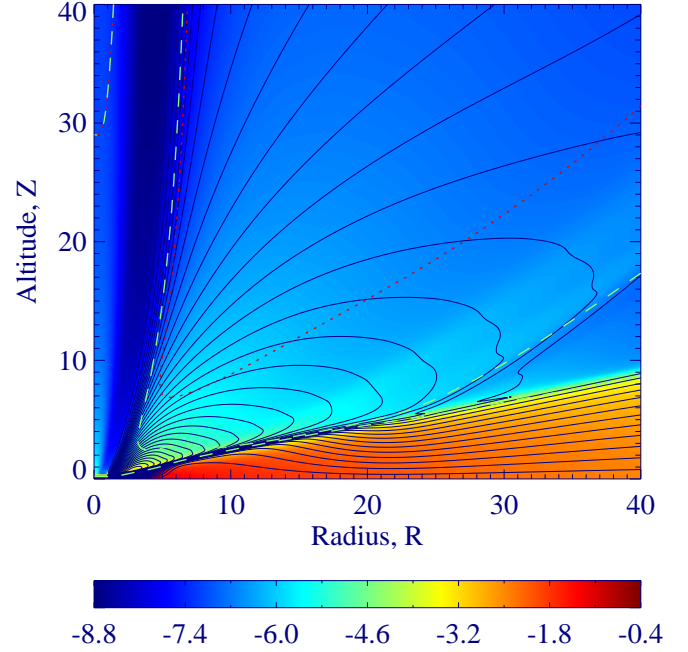}
   \caption{Snapshot at $t=953 \tau_\mathrm{K0}$ of the inner disk regions. The color background is the log of mass density with the Fast (dotted line) and Alfv\'en (dashed line) surfaces overplotted. The last super-FM
   field line is anchored at about $r=5$, whereas the last super-A field line is at $r=13$. The jet exhibits the characteristic butterfly shape in the electric current lines (shown in dark blue).}
   \label{current599}%
\end{figure}

Understanding the behavior of electric currents is the key point in accretion-ejection theory. Figure \ref{current599} shows a zoom of the ejecting regions I and II at $t=953 \tau_\mathrm{K0}$. The Alfv\'en
(dashed) and Fast (dotted line) critical surfaces are shown along with the poloidal electric current lines $\vec j_\mathrm{p}$ (blue). Globally, some current enters the disk at its inner edge ($J_z <0$) and flows
outwardly within the disk. In zones I and II where ejection takes place, current lines are closed within the jet. The crossing of this poloidal current through poloidal field lines results in jet
confinement and acceleration. In zone III where there is no jet, thus much less plasma, there is almost no current flowing there.

\begin{figure*}
   \centering \includegraphics[width=0.9\textwidth]{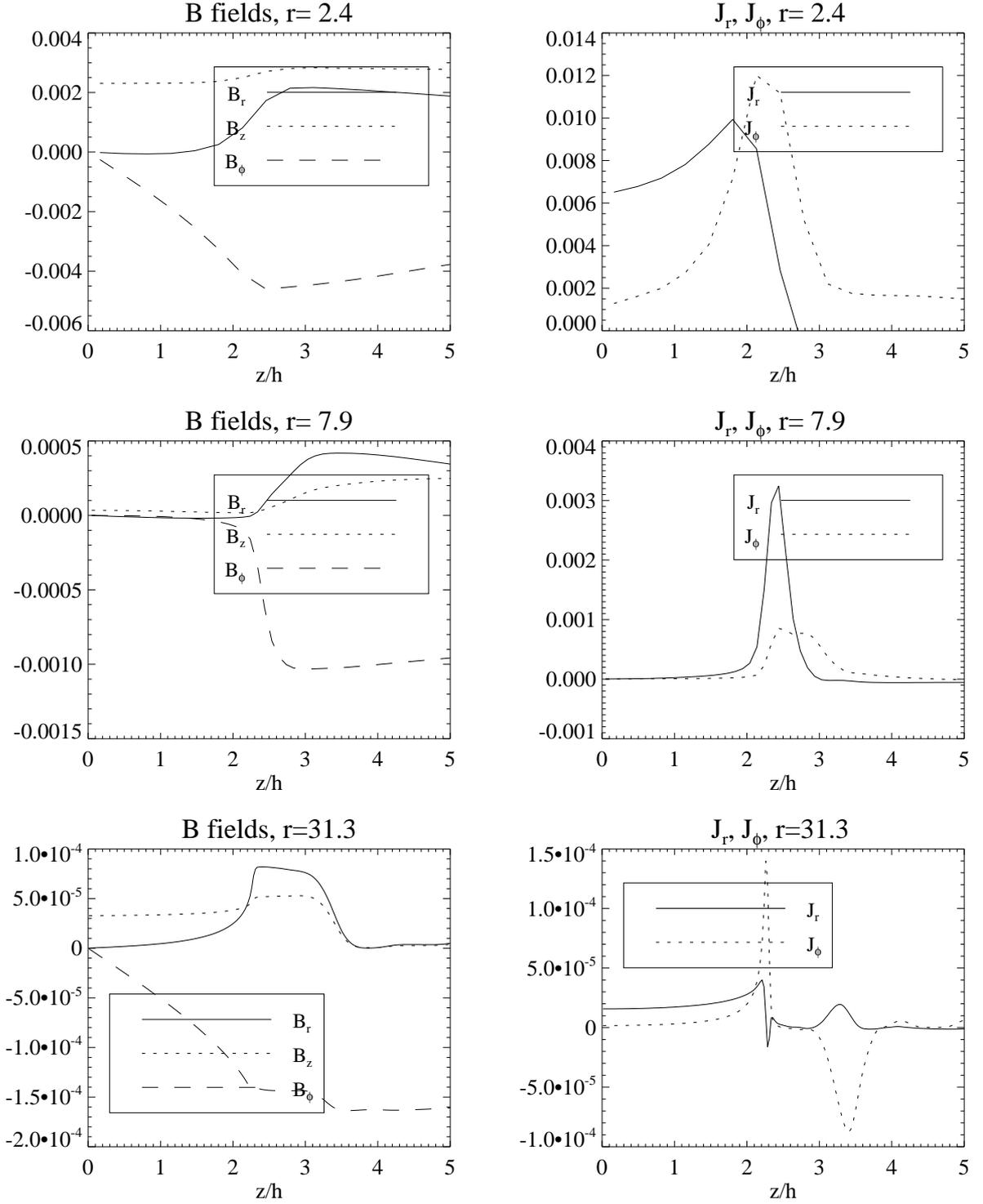}
   \caption{Vertical profiles of the components of the magnetic field (left) and electric current density (right), in zones I ($r=2.4$), II ($r=7.9$) and III ($r=31.3$), plotted at a time
   953$\tau_\mathrm{K0}$. See text for more details.}
   \label{jandb123}%
\end{figure*}

Let us have a look at the vertical profiles of the magnetic field components and electric current density at three radii located within the three previously defined zones (Fig. \ref{jandb123} ). In
all three zones, the disk surface can be easily detected as it is the locus where (i) the radial component undergoes a huge increase and (ii) the toroidal field abruptly changes its behaviour.

Throughout all the disk, the dominant toroidal current density $J_\phi$ is located at the disk surface {\bf not} at the disk midplane as assumed in the infinitely thin disk approximation.  This is
easy to understand from Ohm's law in resistive MHD. Indeed, neglecting the contribution of the vertical velocity one gets
\begin{equation}
J_\phi(z) \simeq - \frac{r u_r}{\nu_\mathrm{m}} \frac{B_z}{r} \; .
\end{equation}
Unless the vertical profile of $u_r$ strictly follows that of the magnetic diffusivity, the vertical decrease of $\nu_\mathrm{m}$ leads unavoidably to an increase of $J_\phi$.  Remarkably, such a profile of
$J_\phi$ has been already discussed in self-similar solutions of \citet{2000A&A...353.1115C} (see their Fig. 7).  As a consequence, field lines remain straight within the bulk of the disk and bend only
at its surface. In all zones, such a bending is large enough to satisfy the Blandford \& Payne energetic criterion for cold wind launching. As a matter of fact, the bending of the field lines
($B^+_r/B_z$) gets larger as one goes from zone I to zone III.

The vertical profile of the toroidal field $B_\phi$ is controlled by the radial current density $J_r$. The profile $J_r(z)$ is very interesting: contrary to self-similar solutions, most of the
poloidal current is flowing at the disk surface and not at the disk midplane. Because of the small value of $B_z$, the unipolar induction effect is small and so is the induced radial current. However,
that current becomes much larger towards the disk surface, mostly because of the vertical decrease of the resistivity. This results in a magnetic shear $|B_\phi^+/B_z |$ that goes from 2.25 (zone I)
to 6.7 (zone II) or even more ($>16$ in zone III). A magnetic shear of $\sim 2$ is a typical value already met in previous simulations and in self-similar models. It results from the interplay between
the disk and the jet and leads to a steady-state. Self-similar models have shown that a larger value will result in unsteady disk and wind configurations
\citep{1993ApJ...410..218W,1997A&A...319..340F}. Remarkably, zone III exhibits an even larger value that still increases with height. There is no jet in this zone but only a torsional Alfv\'en wave
leading to a (negligible) magnetic braking of the underlying disk. In this zone, the magnetic field is so small that the shear can be very large with no actual damage on the disk structure.

In zones I and II, where ejection takes place, the radial current density decreases vertically and becomes eventually negative. In fact a close examination of Fig. \ref{current599} reveals the
following pattern. The poloidal current enters the disk at its surface ($J_z^+<0$) between $r=1$ and $r\simeq2.5$ and then flows inside the disk with $J_r >0$. From $r=2.5$ to about $r=13$ it exits
the disk ($J_z^+ >0$). This is the usual behavior expected in the jet accelerating region, with a typical butterfly shape for the electric poloidal current density \citep[see Fig. 13 in][]{1997A&A...319..340F}. 
This is not the case anymore in zone III where no jet is launched. There, most of the radial current remains confined within the disk, with $J_r^+$ vanishingly small.
This is clearly seen in Fig. \ref{current599}: current lines remain within the disk giving rise to a current $I= 2 \pi r B_\phi$ roughly constant with the radius (thus $B_\phi \propto r^{-1}$). Such
a radial profile of the toroidal field will be discussed in a companion paper.

\subsection{The self-confined jet}
%%%%%%%%%%%%%%%%%%%%%%%%%%%

Here we focus on the dynamics of the super-Fast Magnetosonic outflow referred to as the jet. Such a structure established on a dynamical time scale, namely the local Keplerian time, up to $r=5$. This
is also approximately the time scale for FM waves propagating upstream from the FM surface and reaching the disk surface. As a consequence, the Keplerian time is also the good time scale for
establishing a steady-state. Indeed, we observe that after roughly 30 Keplerian orbits there is no relevant modification of the inner jet structure, corresponding to a few times the orbital period at
$r=5$.

% MHD invariants
  \begin{figure*}
   \sidecaption
     \includegraphics[width=12cm]{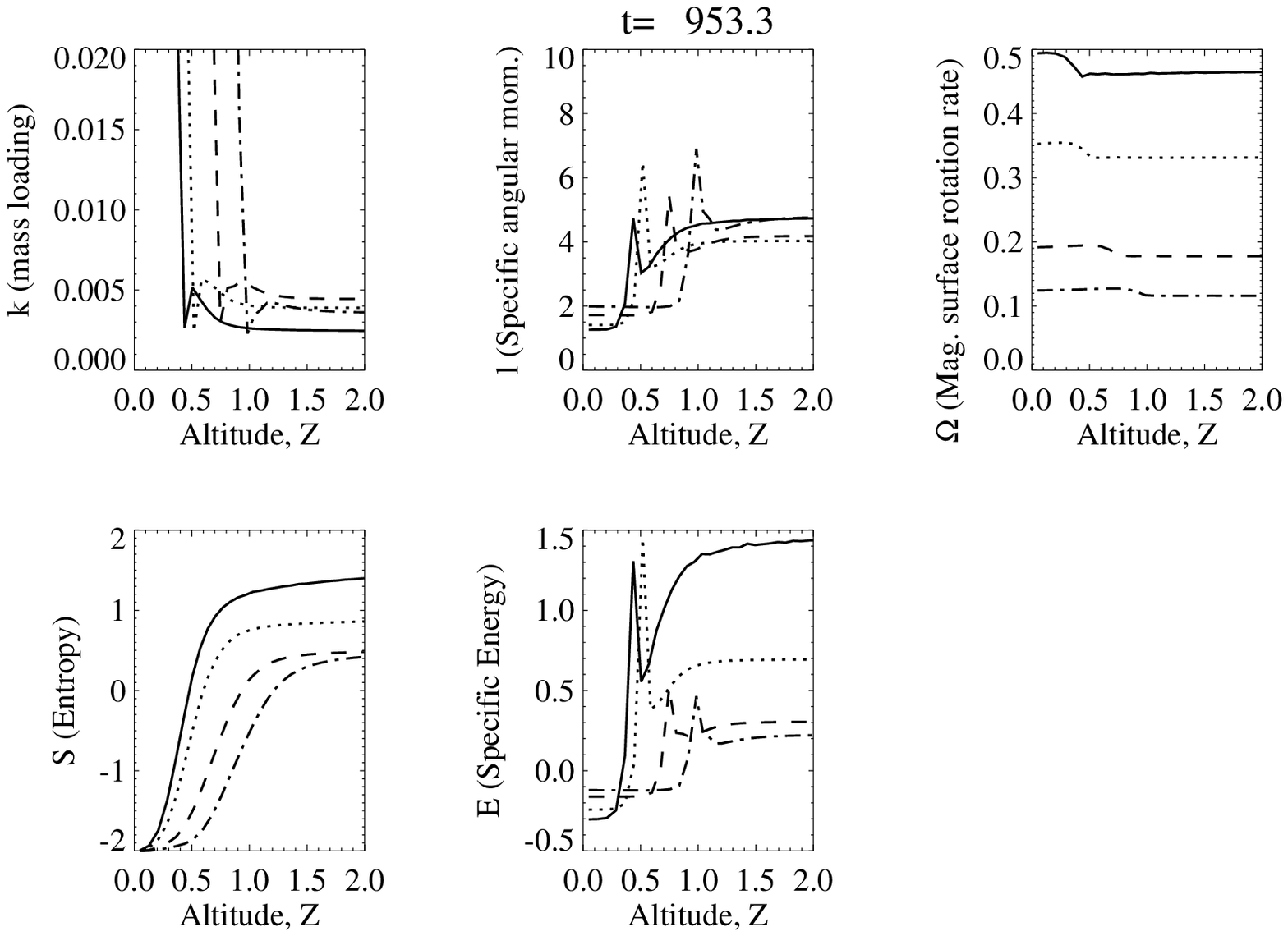}
\caption{The five quantities which are considered invariant under ideal MHD: the mass flux to magnetic flux ratio $k$; the specific angular momentum $l$; the magnetic surface rotation rate
$\Omega_*$; the entropy $S$ and the specific energy or Bernoulli invariant $E$. They are shown in code units along field lines anchored (from left to right in the plot of $S$): $r$ = 1.6, 2, 3, 4. The
measurements are made at $t=953 \tau_\mathrm{K0}$.}
              \label{invar}%
    \end{figure*}

In order to assess whether or not our magnetized adiabatic outflow reached a steady-state, the best way is to compute the five following quantities, namely the mass flux to magnetic flux ratio:
\begin{equation}
k= \rho \frac{u_\mathrm{p}}{B_\mathrm{p}} \; ,
\end{equation}
the specific angular momentum:
\begin{equation}
l= r u_\phi - \frac{r B_\phi}{k} \; ,
\label{eq:l}
\end{equation}
where $u_\phi= \Omega r$, the magnetic surface rotation rate:
\begin{equation}
\Omega_* = \Omega - \frac{k B_\phi}{\rho r} \; ,
\end{equation}
the entropy:
\begin{equation}
S= \log_{10} \left( \frac{P}{\rho^{\gamma}} \right) \; ,
\end{equation}
and the specific energy or Bernoulli invariant:
\begin{equation}
E= \frac{u^2}{2} + \Phi_\mathrm{G} + \frac{\gamma}{\gamma -1} \frac{P}{\rho} - \frac{ \Omega_* r B_\phi}{k} \; .
\label{eq:Ber}
\end{equation}

According to steady state jet theory these quantities should be invariants, namely constant both in time and along each magnetic surface. They are shown along several field lines in Fig. \ref{invar}
as a function of the altitude $z$ at a time 953 $\tau_\mathrm{K0}$.  All quantities first undergo some variation first in the resistive disk region until they become constant. The sudden change in their
profile (see for instance the rise in $E$ and $S$) occurs at the transition from the resistive disk to the ideal MHD flow. Further out, it can be verified that all quantities are indeed invariants,
proving our statement that a steady-state outflow has been achieved.

\begin{figure}
   \centering \includegraphics[width=\columnwidth]{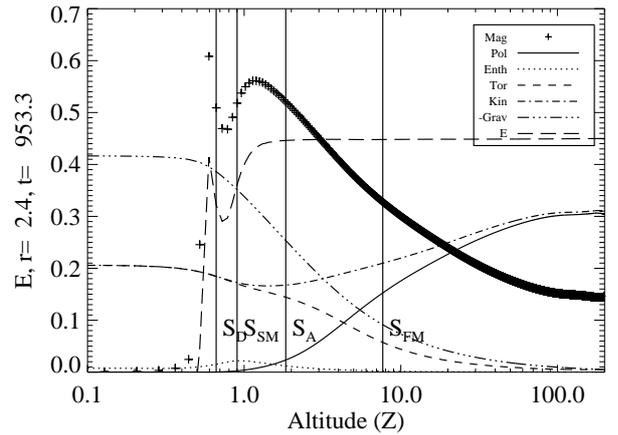}
   \caption{Components of the specific energy $E$ along a magnetic surface anchored at $r=2.4$: magnetic, kinetic (poloidal and toroidal), gravitational and enthalpy. Vertical thick lines indicate the
heights at which the slow magnetosonic (SM), Alfv\'en (A) and fast magnetosonic points (FM) are reached. Notice how negligible is the enthalpy compared with other components.  }
\label{bernoulli}%
    \end{figure}

% The Bernoulli invariant
The specific energy or Bernoulli invariant is separated into its kinetic, enthalpy, magnetic and gravitational components (Eq. \ref{eq:Ber}). These are plotted along a single magnetic surface anchored
at $r=2.4$ in Fig. \ref{bernoulli}.  The vertical line at $z\simeq 0.9$ shows the Slow-Magnetosonic point whereas the second line at $z\simeq 2$ is the Alfv\'en point. The vertical line at $z \simeq
0.7$ shows the altitude where the resistivity has been set to zero, marking thereby the transition between the underlying resistive layers and the ideal MHD flow above. Several important aspects can
be drawn from this plot. First, the enthalpy (solid line) is negligible: we are therefore contemplating a ``cold'' outflow as defined by \citet{1982MNRAS.199..883B}. The large bending of the magnetic
field at the disk surface is thereby enough to drive a magneto-centrifugal outflow. It can be moreover seen that the dominant contribution in $E$ is indeed the magnetic one (+ symbol), as it should be
in such a case. This magnetic energy is then converted into poloidal kinetic energy, but still retains a sizable fraction of its initial value at $z=100$. This limited efficiency of the energy
transfer will be discussed elsewhere.

 It is noteworthy that the specific energy $E$ is an invariant only after roughly $z\sim 1.2$: in all the trans-SM zone, it still increases. This is not due to the enthalpy as it remains always
 negligible despite the huge increase in entropy at the disk-jet interface (see Fig. \ref{invar}). This increase in $E$ can be traced back to the increase in the magnetic component at that same
 location (see the increase in $l$ in Fig. \ref{invar}). This is actually due to a {\em decrease} of the mass to magnetic flux ratio $k$ in the ideal MHD zone. How can this be understood in a well
 tested code where the conservation of quantities such us mass and total energy is assured up to numerical accuracy?

A numerical algorithm such as the one that we employed for our experiments adds to the ``ideal'' flux a diffusive part $F_\mathrm{diff}$ roughly proportional to the local gradient of the corresponding
conserved variable, as in the case of the HLL Riemann solver used in our simulations. For instance, it can be shown that, in a stationary situation, $\vec B_\mathrm{p} \cdot \nabla k$ is proportional to the
divergence of a numerical diffusive flux $F_{\rho, \mathrm{diff}}$ that follows the gradient of the density. As a consequence, our estimator of $k$ is a constant only whenever numerical diffusion is
really negligible.  This is clearly not verified at the disk surface where the steepest gradient is present. But as we move upwards, the numerical contribution vanishes and our estimator converges
(decreases) towards the real value of $k$. This points out however to a possible numerical bias in our MHD simulation.

\begin{figure}
   \centering \includegraphics[width=\columnwidth]{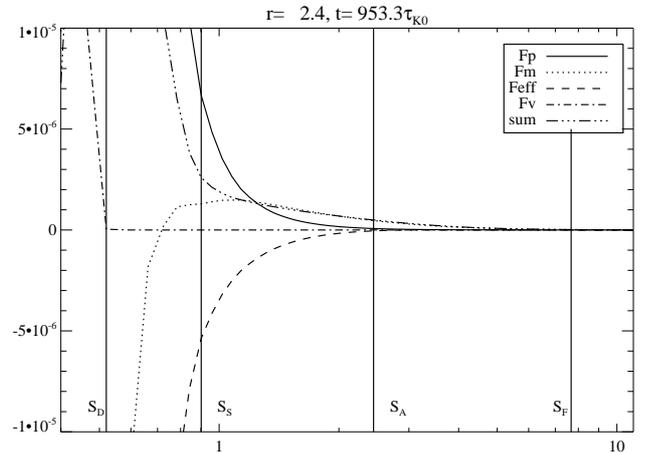}
   \caption{ Forces projected along a poloidal magnetic surface anchored at $r=2.4$, plotted against the altitude above the disk midplane at 953$\tau_\mathrm{K0}$: $F_\mathrm{p}$ is the kinetic pressure gradient,
	$F_\mathrm{m}$ the Lorentz force, $F_\mathrm{eff}$ is the net gravitational+centrifugal forces and $F_\mathrm{v}$ is the viscous stress. The sum of all is also plotted. The vertical lines indicate the heights where
	the flow becomes respectively in ideal MHD (disk surface $S_\mathrm{D}$, super-SM ($S_\mathrm{SM}$), super-A ($S_A$) and super-FM ($S_\mathrm{FM}$).  }
              \label{forceLRes}%
    \end{figure}

%%%%% The jet driving forces
Another numerical bias can be related to irreversible numerical heating, clearly visible in the entropy profiles shown in Fig. \ref{invar} To test this suspicion, let us have a look at the forces that
actually drive the poloidal outflow. It is convenient to compute the projection of all forces along a given magnetic surface (Fig. \ref{forceLRes}).
$F_\mathrm{p}$ is the parallel component of the kinetic pressure gradient
\begin{equation}
F_\mathrm{p} = - \frac{ \vec{B_\mathrm{p}} \cdot \vec \nabla P }{\left| \vec{B_\mathrm{p}} \right|} = - \nabla_\parallel P \; .
\end{equation}
$F_\mathrm{m}$ is the parallel component of the Lorentz force
\begin{equation}
F_\mathrm{m} = \frac{\vec{B_\mathrm{p}} \cdot \left( \vec{J} \times \vec{B} \right) }{\left| \vec{ B_\mathrm{p}} \right|} = - \frac{B_\phi}{2\pi r} \nabla_\parallel I \; ,
\end{equation}
where $I= 2\pi r B_\phi$ is the electric current flowing inside that same magnetic surface \citep{1997A&A...319..340F}, $F_\mathrm{eff}$ is the effective gravitational + centrifugal force
\begin{equation}
F_\mathrm{eff} = \rho \frac{ B_r \Omega^2 r - \vec{B_\mathrm{p}} \cdot \vec \nabla \Phi_\mathrm{G} }{\left| \vec{B_\mathrm{p}} \right|} \; ,
\end{equation}
and, finally, $F_\mathrm{v}$ is the parallel component of the divergence of the shear viscous stress tensor
\begin{equation}
F_\mathrm{v} = \frac{ \vec{B_\mathrm{p}} \cdot \left( \nabla \cdot \vec{T} \right) }{\left| \vec{B_\mathrm{p}} \right|} \; .
\end{equation}
Notice that the centrifugal term contained in $F_\mathrm{eff}$ arises thanks to the azimuthal magnetic force $F_\phi = \left( \vec{J} \times \vec{B} \right) \cdot \vec e_{\phi} = - B_\mathrm{p}
F_\mathrm{m}/B_\phi$. With a magnetic shear $| B_\phi/B_\mathrm{p} |$ of order unity at the base of the jet (Fig. \ref{jandb123}), we have comparable forces $F_\phi \sim F_\mathrm{m} >0$.

What is known from analytical studies is that it is mainly the vertical component of the plasma pressure gradient that lifts the disk material upwards in the resistive MHD layers\footnote{In the zone
where $J_z^+<0$, namely where the current enters the disk surface (from $r=1$ to $r \simeq 2.5$), the toroidal magnetic pressure provides also an upward push ($- \partial_z B^2_\phi >0$). But in the
zone where $J_z^+>0$ (from $r \simeq 2.5$ to $r=5$), the magnetic contribution is only a vertical pinch.}  \citep{1995A&A...295..807F,1997A&A...319..340F}. But this effect works only in a small
vertical extent around the disk surface. This is the region where both the radial and vertical velocity components of the plasma switch from negative (accretion) to positive (ejection). Once this
outward movement has been initiated within the resistive layers, magnetic and centrifugal forces become both dominant and the usual understanding in ideal MHD then applies. The critical issue of mass
loading, namely the amount of mass that is actually ejected from the disk (measured by $k$), is therefore directly related to the delicate interplay of forces in this layer.

This picture is globally confirmed by our numerical simulation (Fig. \ref{forceLRes}). Inside the disk the viscous stress transports momentum upwards but it reduces to zero at the disk surface (see
Appendix \ref{AddNumCond} for the definition).  The projection of the Lorentz force is initially negative, showing that the magnetic force within the disk and up to its surface is hindering ejection,
not helping it. The same holds for the effective force, where gravity overcomes the centrifugal term even up to after the Alfv\'en point. It is indeed the plasma pressure term that makes the
difference by providing a super-SM ejection. Before the Alfv\'en point however, it becomes negligible and the dominant force is the magnetic one $F_\mathrm{m}$.

  \begin{figure}
   \centering \includegraphics[width=\columnwidth]{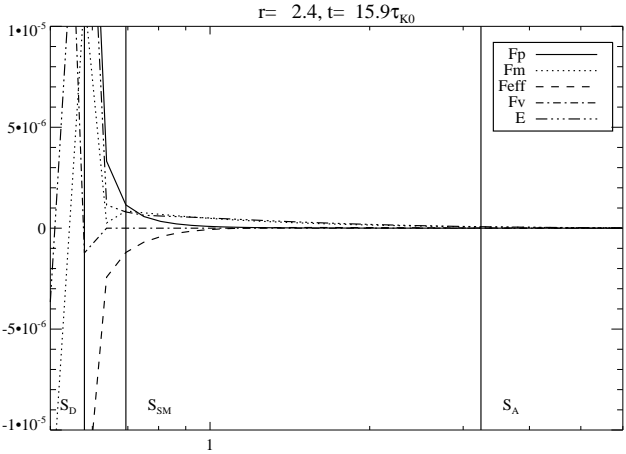}
   \caption{ Same as Fig. \ref{forceLRes} but with a resolution eight times higher. Although the time is now only 15.9$\tau_\mathrm{K0}$, a steady-state has been already achieved at this radius.}
            \label{forceHRes}%
    \end{figure}

This enhanced pressure gradient is likely to be related to the numerical heating visible
in Fig. \ref{invar}. On the other hand, when looking at the Bernoulli invariant, the enthalpy remains always negligible: we argue that, due to the enhanced mass flux related to the diffusive effects
discussed before, the numerical heating per unit volume does not correspond to a significant temperature and enthalpy increase.

\begin{figure*}[t]
\begin{center}
   \begin{minipage}[t]{.48\linewidth}
\includegraphics[width=8cm]{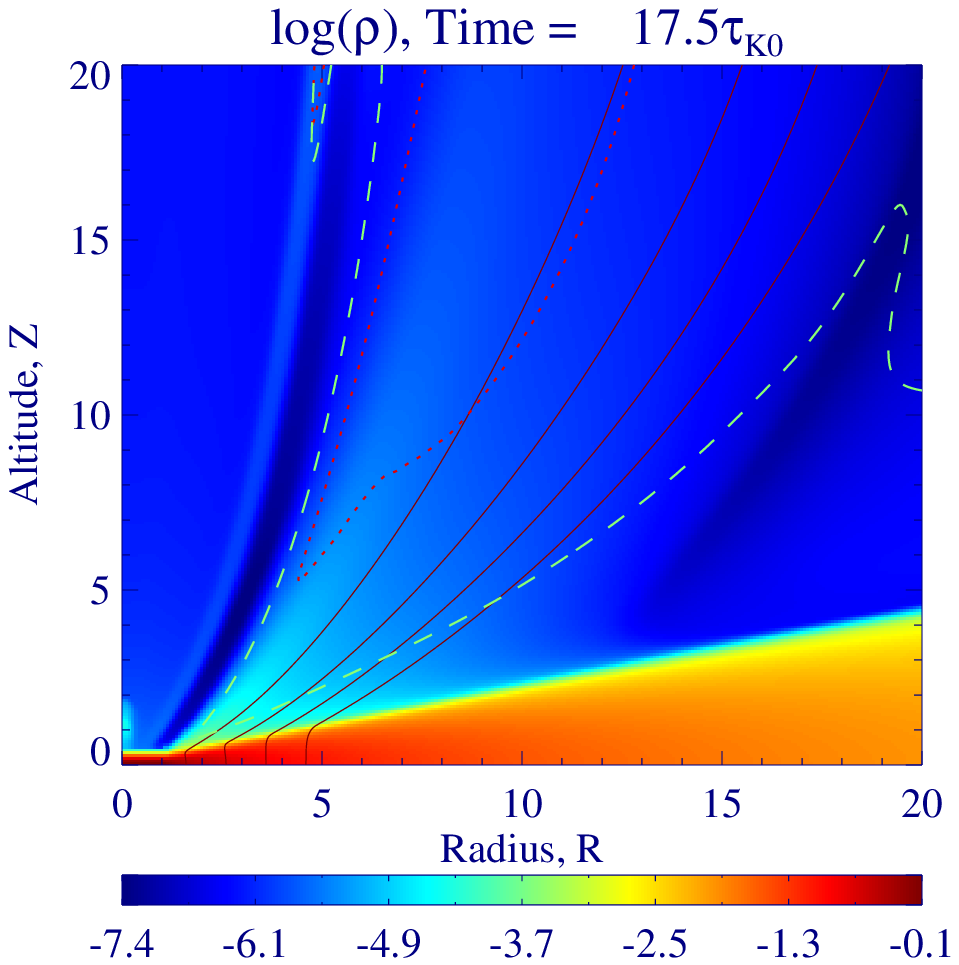}
   \end{minipage} \hfill
   \begin{minipage}[t]{.48\linewidth}
\includegraphics[width=8cm]{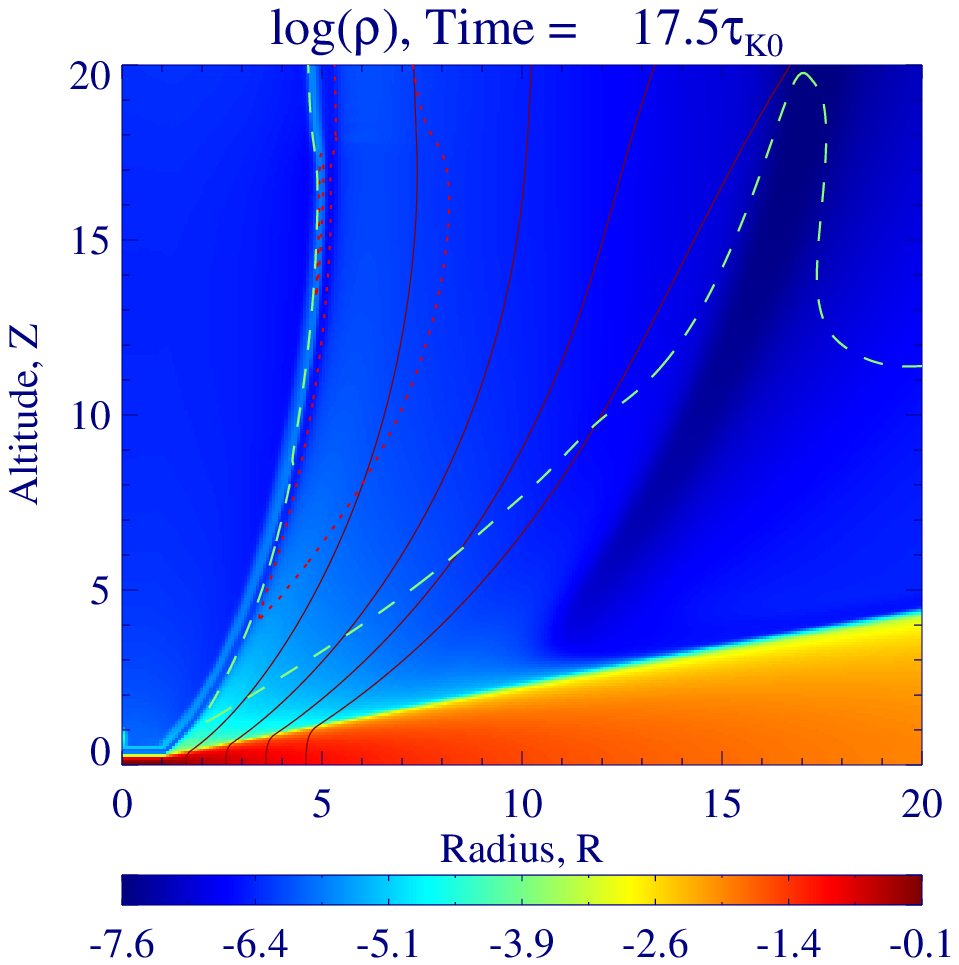}
   \end{minipage} \hfill
\hfill
\caption{ Two simulations done at different resolutions: two times (left panel) and eight times (right panel) our reference simulation. The colormap is the log of gas density with overplotted magnetic
field lines at footpoints $r = 2, 3, 4, 5, 6$. The overlaid critical surfaces are the Alfv\'en surface (dashed line)and the fast magnetosonic surface (dotted line). }
\label{fig:res}
\end{center}
\end{figure*}

\subsection{Mass loading in numerical simulations}
\label{NumericalResolution}
%%%%%%%%%%%%%%%%%%%%%%%%%%%%%%%%%

The side effect of using finite difference methods to solve fluid equations is that it introduces numerical biases that play the role of a magnetic diffusivity, heat conductivity and
viscosity. Although such a numerical diffusion is limited so far as is possible, it is unavoidable and plays a role wherever a steep gradient sets in in any quantity. This is clearly the case for the
density profile (Fig. \ref{denlaunchreg}) at the resistive-ideal MHD zone where the critical issues of mass loading and initial jet acceleration take place. See \citet{2007A&A...469..811Z} for a
discussion of this point. Thus, if present, such a numerical diffusion is actually an extra force term that will appear in particular in the vertical equation and {\em should be present in any
numerical simulation published so far}.

The previous clues that some numerical diffusion of mass is taking place (the bump in the $k$ ``invariant'' seen in Fig. \ref{invar}) can be tested by repeating the simulation at higher resolution. We
therefore performed two more simulations, one with a resolution twice and another with a resolution 8 times higher (Fig. \ref{fig:res}). The physical parameters, boundary and initial conditions
remained unchanged, but the physical extent of the simulation was reduced. Additionally the simulations were only carried out for $\sim 17$ inner disk orbital periods.

\begin{figure*}
   \sidecaption
    \includegraphics[width=12cm]{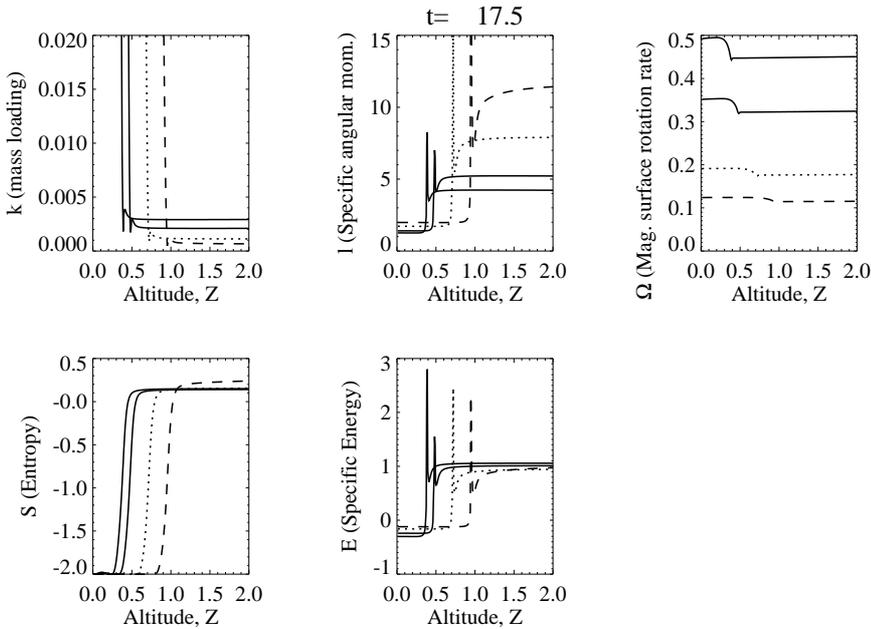}
   \caption{The same as Fig. \ref{invar} but for a resolution eight times higher.  The invariants are flatter and the bump in the mass loading is reduced. A steady-state super-FM jet is still
		present but from a smaller radial extent. The measurements are at time 17.5$\tau_\mathrm{K0}$.  }
              \label{invarhires}%
    \end{figure*}

We obtained the following results : (i) whereas a super-FM jet is still launched in a steady-state, (ii) the radial extent of the ejecting zone is narrower up to $r\simeq 2$ only. We will come to back
to this later. Figure \ref{invarhires} shows the various quantities that should remain invariant, to be compared with Fig. \ref{invar} with the lowest resolution. The anchoring radii are the same as
in Fig. \ref{invar}. Clearly, the invariants are flatter as numerical diffusion is reduced. Also, the bump in $k$ has now almost vanished and the transition from resistive to ideal MHD is much
better caught.  Moreover, the entropy profiles clearly shows that the numerical heating is strongly reduced.

Figure \ref{forceHRes} shows the parallel forces along a given field line anchored at the same radius as in the lowest resolution case. The general trend remains the same although the effect of the
thermal push is now dramatically reduced. It is still the dominant force allowing a trans-SM flow but its importance decreases more rapidly. Of course, only magnetic forces provide a super-Alfv\'enic
flow.
 The reduction of the ejection efficiency with increasing resolution confirms our suspicion: numerical diffusion is indeed at work at the disk surface in the inner regions of the grid. This effect
 naturally explains the mass loading, initial push and thereby increase in the specific energy $E$.

\subsection{The role of the disk magnetization}
%%%%%%%%%%%%%%%%%%%%%%%%%
So far we understood the {\em initial} mass loading and driving mechanisms of the jet. However, what determines its radial extent has remained unexplored. In previous simulations of that kind the
extent of the ejecting zone was increasing in time with a Keplerian scaling $r \propto t^{2/3}$ \citep{2002ApJ...581..988C,2004ApJ...601...90C,2007A&A...469..811Z}. It is the first time where a jet is
launched from a finite region that remains constant over time.

\begin{figure}
   \centering \includegraphics[width=\columnwidth]{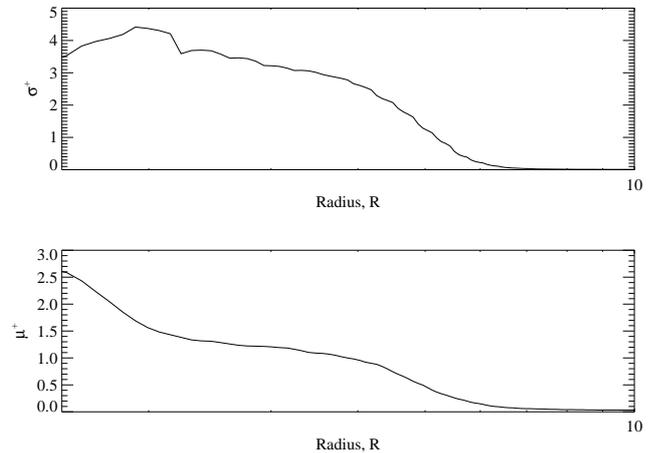}
   \caption{Upper panel: Initial jet magnetization $\sigma^+$ (Poynting to kinetic flux ratio) measured at the disk surface in the inner zones of the accretion disk. Lower panel: Disk magnetization
   ($\mu^+=B_z^2/ P^+$) measured at the disk surface. These curves were obtained with our reference simulation at the final stage.}
              \label{sigmu}%
    \end{figure}

In steady-state jet theory, the Bernoulli invariant must be positive at all magnetic surfaces. Neglecting enthalpy (as Fig. \ref{bernoulli} suggests), Eq. (\ref{eq:Ber}) provides
\begin{equation}
E \simeq \Omega_{*,0}^2 r_0^2 \frac{\sigma^+ - 1}{2} \; ,
\end{equation}
where
\begin{equation}
\sigma^+= - \left . \frac{ 2 \Omega_* r B_\phi B_\mathrm{p} } { \rho u^2 u_\mathrm{p} } \right |_{z=h} \; ,
\end{equation}
is the ratio of the MHD Poynting flux to the kinetic energy flux measured at the disk surface. This quantity is sometimes referred to as the (initial) jet magnetization. A cold jet requires therefore
$\sigma^+$ larger than unity. We plotted in Fig. \ref{sigmu} (top) the jet magnetization as function of the disk launching radius for our reference simulation. Beyond a radius of about 5, this quantity
becomes indeed smaller than unity, corresponding nicely to the end of zone I (super-FM jet). A super-A outflow is nevertheless launched at larger radii, but this is a matter dominated flow that never
reaches a steady-state. The overall picture is therefore consistent. But what determines the radial distribution $\sigma^+(r)$?

Another way to write the initial jet magnetization is
\begin{equation}
\sigma^+= 2 \mu^+ \frac{h}{r} \frac{c_\mathrm{s}}{u_z^+} \left | \frac{B_\phi^+}{B_z} \right | \; ,
\end{equation}
where $\mu^+= B_z^2/P^+$ is measured at the disk surface. We plotted in Fig. \ref{sigmu} (down) the disk magnetization at the upper surface layers. It can be seen that $\sigma^+(r)$ follows
approximately the same trend as $\mu^+(r)$, namely a radial decrease.

 In fact, analytical calculations done within the self-similar framework already pointed out the importance of the disk magnetization $\mu$ for launching super-FM jets. It was shown that isothermal
 \citep{1995A&A...295..807F,1997A&A...319..340F} or adiabatic \citep{2000A&A...353.1115C} magnetic surfaces require a field close to equipartition, namely $\mu$ smaller but around unity. Our own
 results suggest that it is the disk magnetization that actually defines the ejecting zones. Beyond $r=5$, the magnetic field would be too small to allow a proper jet to be launched.

Let us make a very crude approach by assuming a $B_z$ component almost constant in the vertical direction and an isothermal hydrostatic density profile. In that case, one would have $\mu(z) = \mu
\exp(z^2/2h^2)$ where $\mu$ is the disk magnetization at the disk midplane. It seems therefore dubious that $\mu^+$, reaching a value 10 to 100 times $\mu$ at a few scale height, could ever reach a
value of order unity if $\mu$ is too small. So, even in the presence of a numerical diffusion, no jets should be produced if the disk magnetization is too low.

\begin{figure}
   \centering \includegraphics[width=\columnwidth]{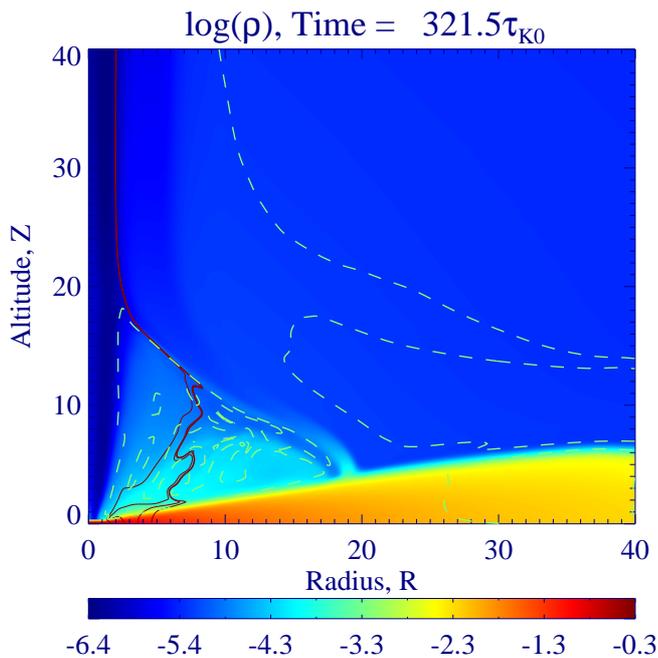}
   \caption{ The colourmap shows the logarithm of gas density in the weak magnetic field simulation at time t=321.5$\tau_\mathrm{K0}$.  The dashed lines enclose material moving at velocities faster than the
	local Alfv\'en speed.  There is no superfast outflow present and the superAlfv\'enic material is extremely fragmented.  }
              \label{weakB}%
    \end{figure}

In order to test this conjecture, we performed another numerical simulation with $\mu$ decreased by one order of magnitude (namely starting at $2\times10^{-4}$ at the disk inner radius). The other
physical parameters, boundary and initial conditions, as well as the numerical resolution were otherwise identical to the reference simulation. We found that, in this case the super-Alfv\'enic
material was extremely sporadic and fragmented in the domain, and no super-fast-magnetosonic jet was observed. See Fig. \ref{weakB}.

In our view, this clearly confirms that the disk magnetization must be high enough in order to launch self-confined (super-FM) jets. This result goes in the same direction as those obtained with
self-similar solutions. However, the latter claimed that only $\mu$ smaller but close to unity (namely a field close to equipartition) allows the launching of magnetized jets. The physical argument is
the following. For a jet to be launched, the lifted mass must cross the SM point around the disk surface. In a cold environment the only force able to do this is the magnetic one. It turns out
however, that it is much easier to do it if the accretion velocity is already not too far from the sound speed. This is the reason why isothermal or adiabatic jets require fields close to
equipartition (Ferreira \& Casse, 2009, submitted).

Apparently, this is in contradiction with our own result since we do obtain jets with $\mu$ of the order of a few $10^{-3}$. The reason for this discrepancy lies in the fact that the analytical models
were obtained under the assumption of either isothermal or adiabatic magnetic surfaces. Here, as we showed, there is a numerical diffusion that allowed mass to leak from the disk to the open, rotating
field lines. This extra effect has been mimicked for instance in \citet{2000A&A...361.1178C} with the presence of a heating term at the disk surface. New solutions, called ``warm'' in contrast to the
previous ``cold'' ones, were found with an enhanced mass flux. But these authors did not recognize that the required disk magnetization $\mu$ was indeed smaller than for cold jets.  We report here
that it is indeed the case, with some of the ``warm'' self-similar solutions found with $\mu=0.08$.

%%%%%%%%%%%%%%%%%%%%%%%%%%%%%%%%%%%%%%%%%%%
\section{Concluding remarks}
\label{ConcludingRemarks}
%%%%%%%%%%%%%%%%%%%%%%%%%%%%%%%%%%%%%%%%%%%

In this paper, we performed four 2.5D numerical MHD simulations of a resistive viscous accretion disk threaded by a weak magnetic field. The initial magnetic field distribution was chosen so that the
disk magnetization $\mu= B_z^2/ P$ decreases radially from the central object. Our reference simulation, done with a maximum value of $\mu= 2\times 10^{-3}$, has run for more than 950 Keplerian orbits
at the inner radius and is therefore the longest to date.

It is shown that the disk structure resembles that of a standard Shakura \& Sunyaev disk with accretion controlled by the turbulent (alpha) viscous torque only. However, a super fast magnetosonic,
self-confined jet is observed to be launched from the inner disk regions. It is first time that (i) steady-state super-FM jets are produced from a weakly magnetized disk and (ii) from a finite disk
region that remained constant over time. The power carried away by these jets is tiny and directly related to the negligible torque on the disk. The dynamics of the jet and its propagation into the
medium will be studied in a forthcoming paper. Here, we focused on the jet acceleration region where the flow crossed the three MHD critical surfaces (Slow Magnetosonic, Alfv\'en and Fast
Magnetosonic).

The critical issues of mass loading and initial jet acceleration (the crossing of the SM surface) are shown to be strongly affected by the unavoidable steep decrease of the density profile at the disk
surface. Such an effect has been underestimated in previous simulations. It is the quality of the grid resolution {\em at the disk surface} that ultimately determines the amount of ejected mass. One
way to solve this problem is to use either an enhanced resolution at the disk surface, a less diffusive algorithm, a higher order method or an adaptive grid which refines on the density gradient.

We argue however that this feature might mimic some additional heat input at the disk surface, as explored for instance by \citet{Ogilvie:1998lh}, \citet{2001ApJ...553..158O},
\citet{2000A&A...361.1178C}.  This aspect is extremely promising as most astrophysical accretion disks do probably have superheated layers due to irradiation by the central object (young stars,
cataclysmic variables) and/or some X-ray source (e.g. around black holes). As a consequence, ``cold'' (e.g. isothermal or adiabatic) ejection is probably never achieved in Nature.

This allows also to relax the constraint of equipartition fields needed for driving jets as our jets were obtained from a very low magnetized disk (but not too low). This opens a new fascinating
topic: the magnetic history of any given object. One might indeed consider accretion disks displaying a whole continuum in ejection efficiency, from jets carrying a sizable fraction (if not most) of
the released accretion power to jets that are a mere epiphenomenon of accretion. For any given object, the key parameter would be the disk magnetization. This clearly deserves further investigation.

\begin{acknowledgements}
This work has been supported by the ANR-05-JC42835 project funded by the ``Agence National de la Recherche'' and through the Marie Curie Research Training Network JETSET (Jet Simulations, Experiments
and Theory) under contract MRTN-CT-2004-005592. The authors wish to acknowledge the SFI/HEA Irish Centre for High-End Computing (ICHEC) for the provision of computational facilities and support.
\end{acknowledgements}

\bibliography{12633}

\begin{appendix}

\section{Additional numerical conditions}
\label{AddNumCond}

As an initial condition for the simulation the perturbative solution of the steady-state MHD equations is taken.  The disk pressure and density are computed by
solving the hydrostatic vertical equilibrium, the toroidal speed is determined by the radial equilibrium, whereas the radial velocity is given by the angular momentum conservation equation.  We
assumed a thermal disk heightscale $h = \epsilon r$, where the aspect ratio $\epsilon = c_\mathrm{s}/V_\mathrm{K} = 0.1$ is given by the ratio between the isothermal soundspeed $c_\mathrm{s} = \sqrt{P/\rho}$ and the Keplerian
speed $V_\mathrm{K} = \sqrt{GM/r}$ calculated at the disk midplane.  The disk density is therefore given by:
\begin{equation}
\rho_\mathrm{d} = \rho_\mathrm{d0} \left\{\frac{2}{5\epsilon^2}\left[\frac{r_0}{R}-\left(1-\frac{5\epsilon^2}{2}\right)\frac{r_0}{r}\right]\right\}^{3/2} \; ,
\end{equation}
and the thermal pressure by:
\begin{equation}
P_\mathrm{d} = P_\mathrm{d0} \left( \frac{\rho_\mathrm{d}}{\rho_\mathrm{d0}} \right)^{5/3} \; ,
\end{equation}
where $P_\mathrm{d0} = \epsilon^2 \rho_\mathrm{d0} V^2_\mathrm{K0}$.  The components of the poloidal speed are:
\begin{equation}
u_{r\mathrm{d}} = -\alpha_\mathrm{v}\epsilon^2\left[10-\frac{32}{3}\Lambda\alpha_\mathrm{v}^2-\Lambda\left(5 -\frac{z^2}{\epsilon^2r^2}\right)\right] \sqrt{\frac{GM}{r}} \; ,
\end{equation}
\begin{equation}
u_{z\mathrm{d}} = u_{r\mathrm{d}} \frac{z}{r} \; .
\end{equation}
The toroidal speed is:
\begin{equation}
u_{\phi\mathrm{d}} = \left[ \sqrt{1-\frac{5\epsilon^2}{2}}+\frac{2}{15}\epsilon^2\alpha^2_\mathrm{v}\Lambda\left(1-\frac{6z^2}{5\epsilon^2r^2} \right) \right] \sqrt{\frac{GM}{r}} \; ,
\end{equation}
with
\[
\Lambda = \frac{11}{5} \big/ \left( 1+\frac{64}{25}\alpha_\mathrm{v}^2\right) \; .
\]
For viscosity and resistivity, the expression used by \citet{2009A&A...508.1117Z} is employed:
\begin{equation}
\nu_\mathrm{v} = \frac{2}{3}\alpha_\mathrm{v} \left[ \left. c^2_\mathrm{s}\left(r\right) \right|_{z=0}+\frac{2}{5}\left( \frac{GM}{R}-\frac{GM}{r}\right)\right]\sqrt{\frac{r^3}{GM}} \; ,
\end{equation}
where the isothermal soundspeed calculated on the disk midplane $\left. c_\mathrm{s}\left(r\right) \right|_{z=0}$ can change in time.

For the atmosphere above disk all velocities are set to zero, $u_r=u_z=u_\phi=0$, and a hydrostatic, spherically symmetric atmosphere is prescribed:
\begin{equation}
\rho=\rho_\mathrm{a0} \left(\frac{r_0}{R}\right)^{\frac{1}{\gamma-1}}
\end{equation}

\begin{equation}
P=\rho_\mathrm{a0} \frac{\gamma-1}{\gamma} \frac{GM}{r_0} \left(\frac{r_0}{R}\right)^{\frac{\gamma}{\gamma-1}} \; .
\end{equation}

A density contrast $\rho_\mathrm{a0}/\rho_\mathrm{d0} =10^{-4}$ has been assumed in all the simulations.

The components of the viscous stress tensor ${\overline {\overline {\vec T}}}$ used in PLUTO are:
% % %%%%%%%

\begin{equation}
T_{rr}=2 \eta_\mathrm{v} \frac{\partial u_r}{\partial r} + \left(\zeta - \frac{2}{3}\eta_\mathrm{v} \right) \vec \nabla \cdot \vec{u}
\end{equation}

\begin{equation}
T_{zz}=2 \eta_\mathrm{v} \frac{\partial u_z}{\partial z} + \left(\zeta - \frac{2}{3}\eta_\mathrm{v} \right) \vec \nabla \cdot \vec{u}
\end{equation}

\begin{equation}
T_{\phi\phi}=2 \eta_\mathrm{v} \frac{ u_r}{ r} + \left( \zeta - \frac{2}{3}\eta_\mathrm{v} \right) \vec \nabla \cdot \vec{u}
\end{equation}

\begin{equation}
T_{\phi r} \equiv T_{r\phi}= \eta_\mathrm{v} \left( \frac{\partial u_\phi}{\partial r} -\frac{u_{\phi}}{r} \right)
\end{equation}

\begin{equation}
T_{zr} \equiv T_{rz}= \eta_\mathrm{v} \left( \frac{\partial u_z}{\partial r} + \frac{\partial u_r}{\partial z} \right)
\end{equation}

\begin{equation}
 T_{\phi z} \equiv T_{z\phi}= \eta_\mathrm{v} \left( \frac{\partial u_\phi}{\partial z} \right) \; ,
\end{equation}
where the bulk viscosity $\zeta$ is set to zero.

\end{appendix}
\end{document}